\documentclass[preprint,10pt]{elsarticle}

\usepackage{graphicx,bbm,setspace,amssymb,amsfonts,amsmath,mathtools,mathrsfs,stmaryrd,amsthm,bm,etoolbox,comment,hyperref,url,caption,subcaption,color,enumitem,algorithm,booktabs,microtype,nicefrac,lingmacros,nccmath,tree-dvips,tikz-cd}
\usepackage[noend]{algpseudocode}
\usepackage{algorithm}

\usepackage[colorinlistoftodos]{todonotes}
\usepackage[utf8]{inputenc} 
\usepackage[T1]{fontenc}    

\DeclareMathOperator*{\argmax}{argmax}
\DeclareMathOperator*{\argmin}{argmin}
\usepackage[noend]{algpseudocode}

\newtheorem{thm-defn}[theorem]{Theorem/Definition}

\theoremstyle{definition}
\newtheorem{theorem1}{Theorem}[section]
\newtheorem{definition}[theorem1]{Definition}

\theoremstyle{remark}

\DeclareMathOperator{\E}{\mathbb{E}}

\newcommand{\ignore}[1]{}{}

\begin{document}

\begin{frontmatter}

\title{Temporal and spectral governing dynamics of Australian hydrological streamflow time series}
   
\author[label1]{Nick James} \ead{nick.james@unimelb.edu.au}
\author[label1]{Howard Bondell} 
\address[label1]{School of Mathematics and Statistics, University of Melbourne, Victoria, Australia}

\begin{abstract}
We use new and established methodologies in multivariate time series analysis to study the dynamics of 414 Australian hydrological stations' streamflow. First, we analyze our collection of time series in the temporal domain, and compare the similarity in hydrological stations' candidate trajectories. Then, we introduce a Whittle Likelihood-based optimization framework to study the collective similarity in periodic phenomena among our collection of stations. Having identified noteworthy similarity in the temporal and spectral domains, we introduce an algorithmic procedure to estimate a governing hydrological streamflow process across Australia. To determine the stability of such behaviours over time, we conclude by studying the evolution of the governing dynamics and underlying time series with time-varying applications of principal components analysis (PCA) and eigenspectrum analysis. 

\end{abstract}


\end{frontmatter}

\section{Introduction}
\label{Introduction}

With the recent focus on long-term shifts in temperature and weather patterns, and the potentially devastating impact for the future, the importance of earth and environmental science may never have been greater. There is a need for continued and varied studies of country hydrology behaviours, to ensure features such as streamflow are not exhibiting significant changes in their behaviour. This paper explores streamflow patterns among Australian hydrology stations between 01 January 1980 and 01 January 2019.

Our work builds upon a substantial body of \emph{multivariate time series analysis} in applied mathematics and statistics, which has been used in a variety of domains including epidemiology,\cite{james2021_CovidIndia, james2021_TVO,James2021_geodesicWasserstein,james2020covidusa,Chowell2016,jamescovideu,Manchein2020,Blasius2020,James2021_virulence} financial markets,\cite{arjun, james2021_crypto2, james2021_MJW, Drod2020_entropy,Jamesfincovid,james2021_mobility, JamescryptoEq} and other fields such as epidemiology and human behaviours.\cite{Vazquez2006,Mendes2018,Shang2020, james2021_olympics} Such techniques that have been used are broad and include parametric models, \cite{Hethcote2000,Perc2020} various aspects of distance and similarity measures, \cite{Moeckel1997,Szkely2007,Mendes2019,James2020_nsm} networks, \cite{Karaivanov2020,Ge2020,Xue2020} clustering frameworks \cite{Machado2020} and various statistical learning settings. \cite{Ngonghala2020,Cavataio2021,Nraigh2020,Glass2020}

There is a significant collection of work studying spatio-temporal patterns among hydrological processes \cite{mauget_multidecadal_2003, sanborn_predicting_2006, amrit_relationship_2018}. Although the scope of such studies varies, there is generally a consistent theme of exploring the change in temporal behaviours at different locations in space. One feature of particular interest to hydrology researchers has been streamflow modelling \cite{redmond_surface_1991, bales_identification_2001}, where a significant amount of work has focused on modelling regional locations across the United States  \cite{cayan_enso_1999, hamlet_long-range_2000, hidalgo_enso_2003, nowak_colorado_2012, wise_hydroclimatology_2018}. Furthermore, recent work has explored the use of wavelets in studying time series predictability within this domain \cite{Guntu2020}. Many of these studies have taken a more concentrated approach, providing detailed analysis on relatively few hydrology stations. This paper takes a different approach, studying more than 400 stations, sampled from a broad range of geographic locations across Australia. 

A great deal of previous research in the hydrology community has used traditional linear and nonlinear statistical modelling techniques such as linear regression \cite{liu_enso_2001}, multivariate linear regression and self-organizing maps (SOM) \cite{barros_toward_2008} and spatial modelling techniques \cite{sun_multiple_2012} to yield insights related to topics such as El Ni\~{n}o Southern Oscillation (ENSO), drought, hydroclimatic variability and streamflow. In more recent years, many researchers have exploited advances in statistical learning, using various artificial intelligence methods such as support vector machines \cite{fung_improved_2020, maity_potential_2010} and deep learning \cite{kaur_deep_2020, khan_prediction_2020} to address problems related to prediction and assessment. These methods are often criticised for the difficulty one has in interpreting the drivers behind their predictions. 
In terms of specific applications of (relatively) recent work, some examples include Dickinson et al. and Najibi et al. \cite{Dickinson2019,Najibi2017} who study the seasonality of climatic drivers behind flood variability and drivers behind long-duration floods, respectively. Similarly Ryberg et al. \cite{Ryberg2016} explore tree-ring based estimates of long-term seasonal precipitation, while Whitfield et al. \cite{Whitfield2020} explore spatial patterns of temporal changes in streamflow data. Our work asks similarly-themed questions and uses recently established and long-standing techniques in time series analysis to study temporal, spectral and spatial dynamics related to Australian streamflow, and subsequently explore collective similarity among our collection of hydrology stations. 

Although the study of nonstationary processes \cite{Dahlhaus1997, Adak1998}, and the application of spectral analysis have been widely studied in the statistics community \cite{Kohn1997, Wood2017, HadjAmar2019, james2021_spectral, Choudhuri2004}, there have been relatively few studies applying such techniques to address problems in hydrology. Notable studies of stationarity include the exploration of spatial correlations between daily streamflows \cite{betterle_what_2017}, monthly streamflow forecasts \cite{gibbs_state_2018}, annual streamflow and flood peaks \cite{milly_stationarity_2008}. Spectral analysis has been applied in hydrology to address several specific problems including the study of watershed properties \cite{Schuite2019} and aquifiers \cite{Gelhar1974, Jimenez2013, Pedretti2016}. This paper uses a variety of techniques for the study of nonstationary processes and time-varying power spectra from the statistics literature, and applies them to model periodic phenomena in Australian streamflow. Furthermore, we explore the evolution of spectral patterns - testing for shifts in periodic dynamics over time. 

To test for changes in behaviour over time, we explicitly, and non-parametrically, model the evolutionary dynamics of our underlying time series. Such studies of time-varying dynamics, correlation structure and dimensionality reduction have been widely applied among applied mathematicians in domains such as finance \cite{Fenn2011, Laloux1999, Mnnix2012, Laloux1999, Kim2005, Pan2007, Wilcox2007}. In hydrological research however, there is less abundance in the applications of such techniques \cite{barros_toward_2008}. We build upon studies of time-varying correlation structure to generate insights into collective temporal and spectral behaviours in Australian streamflow.

Throughout this paper, we refer to a governing physical hydrological process. In this paper, we refer to the underlying dynamics governing Australian streamflow data evolving over space and time. In other hydrological contexts physical processes may refer to precipitation, energy balance, hydraulics, infiltration, etc. Throughout this paper, we focus on streamflow data. 

This paper is structured as follows. In Section \ref{Data} we describe the data used in this paper. In Section \ref{Time_domain_analysis} we study the collective similarity of our underlying stations in the time domain. In Section \ref{Spectral_domain_analysis} we apply an optimization framework and spectral analysis methodology to determine the similarity in various hydrological stations' periodic behaviours. In Section \ref{Governing_physical_process} we introduce an algorithmic procedure to estimate a governing physical process, from the underlying stations. In Section \ref{Evolutionary_Dynamics} we explore the evolutionary dynamics in the governing physical process and the underlying time series, using a variety of techniques such as PCA and associated eigenspectrum analysis implemented in a time-varying capacity. In Section \ref{conclusion} we conclude.

\section{Data}
\label{Data}
Our dataset consists of Australian hydrology stations, their location, catchment area and streamflow and is sourced from the Australian Bureau of Meteorology (http://www.bom.gov.au/water/hrs). We only study stations with reported daily streamflow (ML/day) between 01-01-1980 - 01-01-2019. This provides a total of 414 stations to be studied throughout this paper. 

\section{Temporal dynamics and similarity}
\label{Time_domain_analysis}

\begin{figure*}
    \centering
    \begin{subfigure}[b]{0.48\textwidth}
        \includegraphics[width=\textwidth]{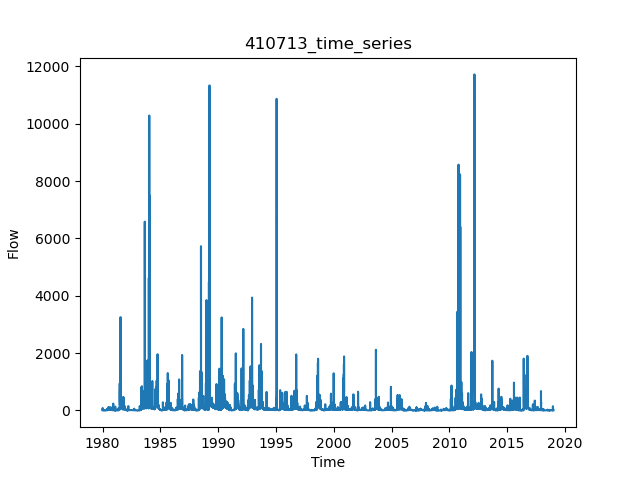}
        \caption{Paddy's River at Riverlea}
    \label{fig:Time_1}
    \end{subfigure}
    \begin{subfigure}[b]{0.48\textwidth}
        \includegraphics[width=\textwidth]{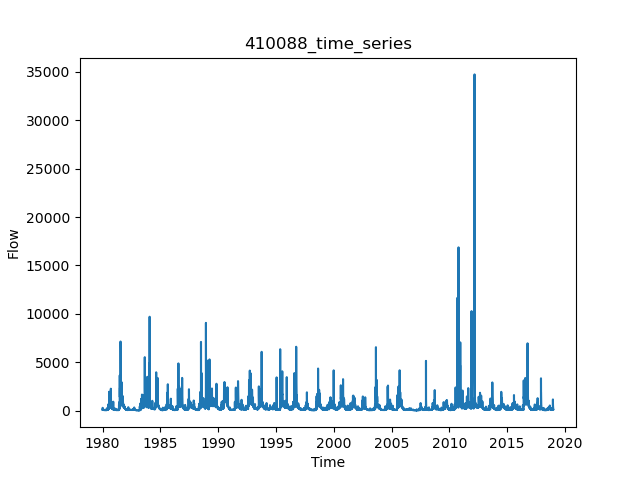}
        \caption{Goodradigbee River at Brindabella}
    \label{fig:Time_2}
    \end{subfigure}
    \begin{subfigure}[b]{0.48\textwidth}
        \includegraphics[width=\textwidth]{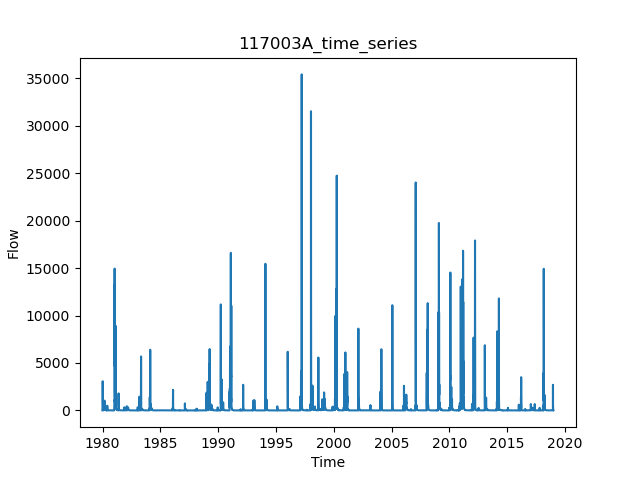}
        \caption{Bluewater Creek at Bluewater}
    \label{fig:Time_3}
    \end{subfigure}
    \begin{subfigure}[b]{0.48\textwidth}
        \includegraphics[width=\textwidth]{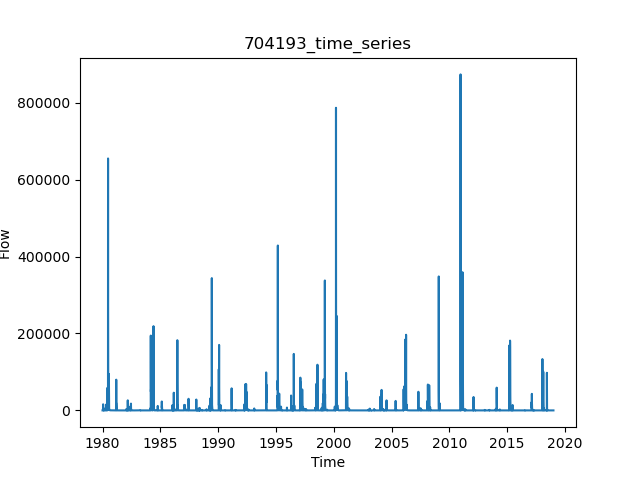}
        \caption{Gascoyne River at Fishy Pool}
    \label{fig:Time_4}
    \end{subfigure}
    \caption{4 candidate flow time series from various locations in Australia. Each location clearly displays highly periodic behaviour.}
    \label{fig:Time_hydrological_time_series}
\end{figure*}

Our central object of study throughout this paper is a collection $n=414$ streamflow time series indexed $i=1,...,n$. We study daily data across $T = 14,246$ days. We let $x_i(t) \in \mathbb{R}$ be the multivariate time series of daily streamflow of station $i$ on day $t$. Throughout this section, we study the collective temporal similarity between all hydrological stations during our window of analysis. First, we visualise all the time series and inspect notably persistent trends across all stations. Figure \ref{fig:Time_hydrological_time_series} displays 4 candidate time series from our collection, sampled from the Eastern, Northern and Western spatial extremes of Australia (four different states) to maximize potential spatial variability in their underlying behaviours. Figure \ref{fig:Time_1} corresponds to the Paddy's River at Riverlea station, situated in The Australian Capital Territory with latitude -35.3843$^{\circ}$ and longitude 148.9656$^{\circ}$. Figure \ref{fig:Time_2} displays streamflow at the Goodradigbee River at Brindabella station, which is situation in New South Wales with latitude -35.42$^{\circ}$ and longitude 148.73$^{\circ}$. Figure \ref{fig:Time_3} displays the Bluewater Creek station at Bluewater, situated in Queensland with latitude -19.185$^{\circ}$ and longitude 146.5461$^{\circ}$. Finally, Figure \ref{fig:Time_4} displays daily streamflow for the Gascoyne River at Fishy Pool station, which is situated in Western Australia and located at latitude -24.9497$^{\circ}$ and longitude 114.6442$^{\circ}$. Despite their spatial variability, all streamflow time series present several similar features.

First, they all possess a similar annual spike in their streamflow values. Second, all locations display a similarly subdued level of streamflow prior to periodic spikes. Such periodic spikes appear to occur at similar times of the year, although this alignment between time series appears to deviate up to a small degree of perturbation. One key difference between the 4 stations visualised is the amplitude of their respective signals, which exhibit pronounced differences and can be seen by the upper bound of the y-axis. For instance, the peak daily streamflow in Paddy's River at Riverlea is approximately 12,000 ML/day while the Gascoyne River at Fishy Pool is in excess of 800,000 ML/day.


To further investigate the structure of temporal similarity throughout our collection, we normalize each streamflow time series by its total amplitude and compute distances between the resulting time series. To do so, let $||x_i||=\sum_{t=1}^T {x}_i(t)$ be the $L^1$ norm of the vector $x_i(t)$. As all ${x}_i(t)$ are non-negative, this aggregates the total amount of streamflow observed between 1980-2019. We then define $\tilde{x}_i(t)= \frac{x_i(t)}{||x_i||}$. The vectors $\tilde{x}_i(t)$ reflect relative changes in streamflow, and capture periods of relatively more activity having normalised for amplitude. Using our earlier figures for the sake of exposition, a change from 0 to 10,000 in Figure \ref{fig:Time_1} would constitute a more significant change than a change from 600,000 to 800,000 in Figure \ref{fig:Time_4}.
We now define a \emph{temporal distance matrix} $D^{T}_{ij}= \sum^T_{t=1} |\tilde{x}_i(t) - \tilde{x}_j(t)|$ that measures distance between  normalized trajectories. To better interpret inherent similarity, we associate a $n \times n$ affinity matrix $A^{T}$ by
\begin{equation}
A^{T}_{ij} = 1 - \frac{D_{ij}^{T}}{\max D^{T} },   
\end{equation}
Our resulting \emph{temporal affinity matrix} $A^{T} \in \mathbb{R}^{n \times n}$ is normalised such that all matrix elements now lie in $[0,1]$. We apply hierarchical clustering to the resulting affinity matrix, and study the collective temporal similarity. 

\begin{figure*}
    \centering
    \includegraphics[width=0.95\textwidth]{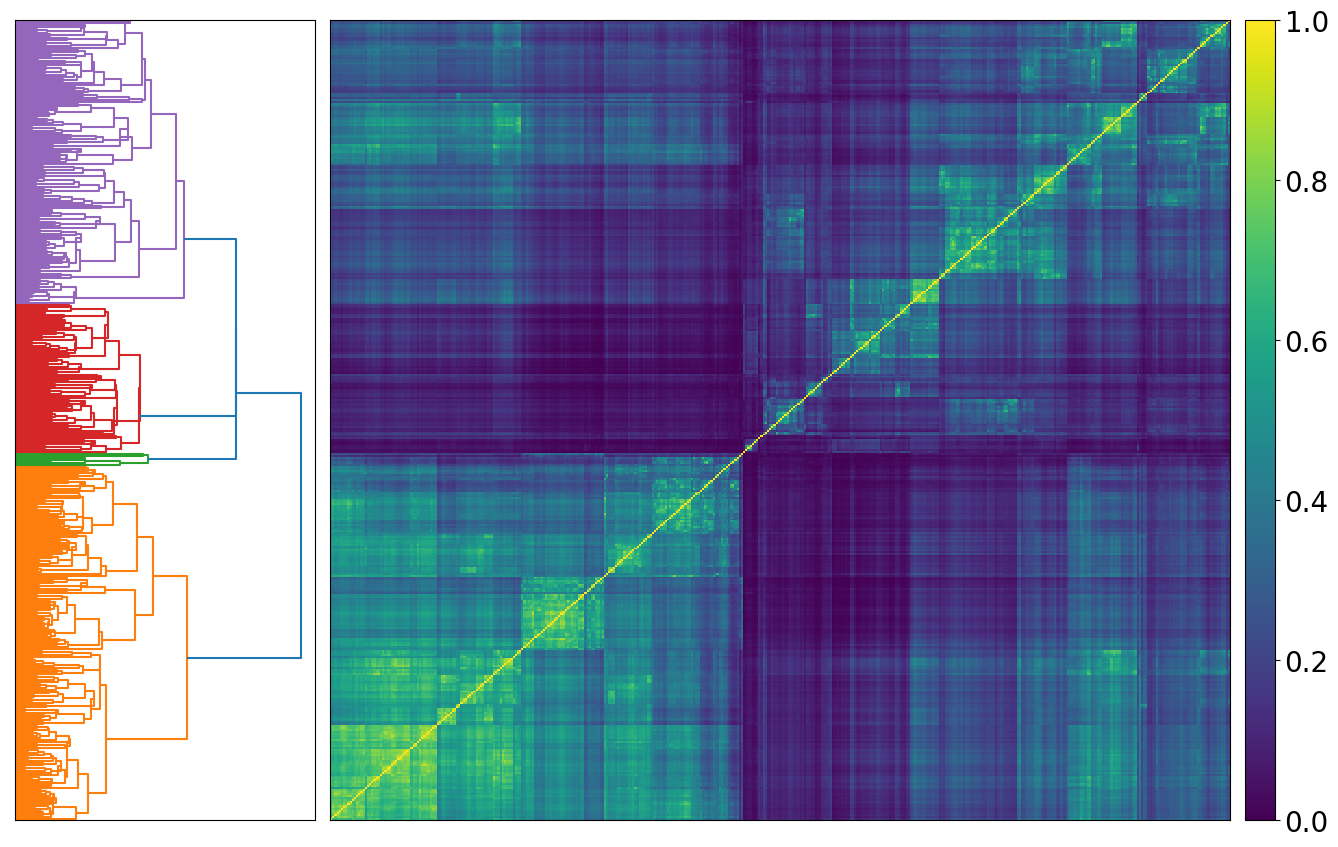}
    \caption{Temporal trajectory distance matrix applied to flow time series. One can identify three predominant clusters (and a small outlier cluster) marked in purple, red, orange (and green) respectively. The red cluster displays a high degree of self-similarity relative to other clusters.}
    \label{fig:Dendrogram_temporal_trajectories}
\end{figure*}

Hierarchical clustering on the temporal affinity matrix is shown in Figure \ref{fig:Dendrogram_temporal_trajectories}. The mean level of temporal similarity is determined by summing all matrix elements above the diagonal and normalizing appropriately. That is, $ \nu^{T} = \frac{2}{n \times (n-1)} \sum_{1 \leq i < j \leq N} A_{ij}^{T} $. We refer to this as an affinity trajectory norm $\nu^T = 0.26$. However close inspection of the associated dendrogram reveals noteworthy distinction between various inter-cluster and intra-cluster similarities. Four clusters of temporal streamflow time series are identified, three of which are most prominent. The two largest clusters, displayed at the top (purple) and bottom (orange) of the dendrogram both exhibit high levels of intra-cluster similarity, and although not quite as high, reasonably high affinity between each other. The third primary cluster which is shown in red, displays markedly lower levels of similarity when compared with constituents of the purple and orange clusters. Having inspected the location of individual stations, cluster membership does not appear to be related to spatial location, but rather, temporal dynamics that are difficult to detect.


\section{Spectral dynamics and similarity}
\label{Spectral_domain_analysis}
Given the pronounced similarity in the stations' spikes in streamflow, we wish to further explore the periodic nature of this phenomenon. Spectral analysis is the most prolific technique in the physical sciences for understanding a time series' autocovariance structure. The most important properties to study are the peak amplitude and respective frequencies of a candidate power spectrum. In this section, we propose a Whittle \cite{Whittle1954, Whittle1957} Likelihood-based optimization framework, to learn the most appropriate parameters used in Welch's method \cite{Welch1967} for power spectral density estimation. In doing so, we are able to model the spectral dynamics of our entire collection of streamflow spectra. In describing our spectral analysis model for streamflow time series, we drop the subscript $i$, referencing specific time series, for notational convenience.

\subsection{Spectral analysis model}

We commence our modelling procedure with our underlying streamflow time series, which we denote $x(t)$. 
\begin{definition}
$x(t)$ is deemed to be stationary if: 
\begin{enumerate}
    \item Each random variable $x(t)$ is integrable with a finite mean $\mu$ for each point in time $t$. We detrend our time series, and henceforth assume that we have a process where $\mu=0$.
    \item The underlying autocovariance  $\E[(x(t)-\mu)(x(t+k)-\mu)]$ is only a function of $k$, which we denote $\gamma(k)$.
\end{enumerate}
\end{definition}
In our application of spectral analysis, we want to study the second-order properties of a stationary time series expressed in the time series' autocovariance. The power spectral density of such a time series is defined: 
 \begin{equation}
     f(\nu) = \sum^{\infty}_{k=-\infty} \gamma (k) \exp (-2\pi i\nu k), \text{ for} -\frac{1}{2} \leq \nu \leq \frac{1}{2}.
 \end{equation}
 
Using frequency components which are termed Fourier frequencies $\nu_j=\frac{j}{n}, j=0,1,...,n-1$, we define the \emph{Discrete Fourier transform} of our streamflow time series:
\begin{equation}
Z(\nu_j) =\frac{1}{\sqrt{n}} \sum_{t=1}^n x(t) \exp(-2 \pi i \nu_j t), \text{ for } j = 0, ... ,n-1.
\end{equation}

For the $j^{th}$ Fourier frequency component, $\nu_{j}$, an unbiased (noisy) estimate of the power spectral density at that point, $f(\nu_{j})$, is provided by the \emph{periodogram}, which is defined by $I(\nu_{j})=|Z(\nu_j)|^2$. By symmetry, the periodogram contains $m=[\frac{n}{2}]+1$ effective observations, which corresponds to  $j=0,1,...,m-1$. Previous work has provided a parsimonious signal plus noise representation of the periodogram:
\begin{equation}
\label{eq:log_periodigram}
\log I(\nu_{j}) = \log f(\nu_{j}) + \epsilon_{j}, \text{ where } \epsilon_{j} \sim \log(\exp(1)),
\end{equation}
which we will use in our subsequent analysis. Many approaches, especially in the Bayesian statistics community, use the Whittle likelihood function to non-parametrically model the log periodogram within a regression framework. The likelihood of the log periodogram conditional on the true power spectrum can be approximated by
\begin{equation}
p(\log I(\nu_j)| \log f(\nu_j)) = (2 \pi)^{-m/2} \prod_{j=0}^{m-1} \exp\left({-\frac{1}{2}\left[\log f(\nu_{j}) + \frac{I(\nu_{j})}{f(\nu_{j})}\right]}\right).
\end{equation}

We use Welch's method to yield an estimate of the latent power spectral density. Welch's method aims to reduce noise in its estimating the power spectrum by sacrificing lower frequency resolution. Procedurally, the data is divided into overlapping segments where a modified periodogram is computed within each partition. The modified periodograms are averaged to produce a final estimate of the power spectrum. Welch's method consists of two model parameters: the length of each segment, and the amount of overlap in data points between adjacent segments. Overlap is often defined in percentage terms, where 0\% would mean that two adjacent segments have no data overlap while 50\% would mean that any segment would share half the data of a neighbouring segment.


Let our collection of power spectral density estimates yielded by Welch's method be $f_i^{W}(\nu_{j})$, for all hydrological stations $i=1,...,n$. Such estimates are highly dependent on model parameters $\Theta = \{S, \omega \}$, where $S$ is the size of each segment and $\omega$ is the overlap between neighbouring segments. To best estimate our optimal power spectrum for each streamflow time series, we determine parameter values which optimize the Whittle Likelihood over our entire collection of time series. This optimization can be written
\begin{equation}
    \argmax_{\Theta} \sum^n_{i=1} \sum^{m-1}_{j=0} \log f^{W,(i)}_{\Theta} (\nu_j) + \frac{I^{(i)}_{\Theta} (\nu_j)} {f^{W,(i)}_{\Theta} (\nu_j)}.
\end{equation}

For this data, the above optimization results in parameter estimates of $S^{*} = 3750$ and $\omega^{*} = 0.4$. We apply Welch's method using these optimal parameters to our streamflow time series, and generate our collection of \emph{optimal spectra} $f_i^{*}(\nu_j)$. 

\begin{figure*}
    \centering
    \begin{subfigure}[b]{0.48\textwidth}
        \includegraphics[width=\textwidth]{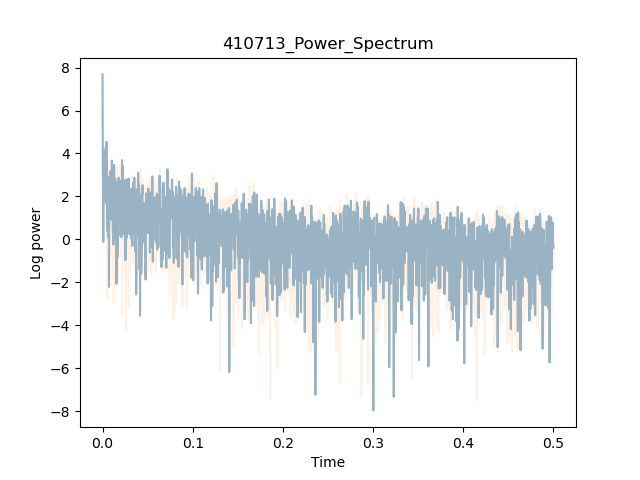}
        \caption{}
    \label{fig:Spectral_1}
    \end{subfigure}
    \begin{subfigure}[b]{0.48\textwidth}
        \includegraphics[width=\textwidth]{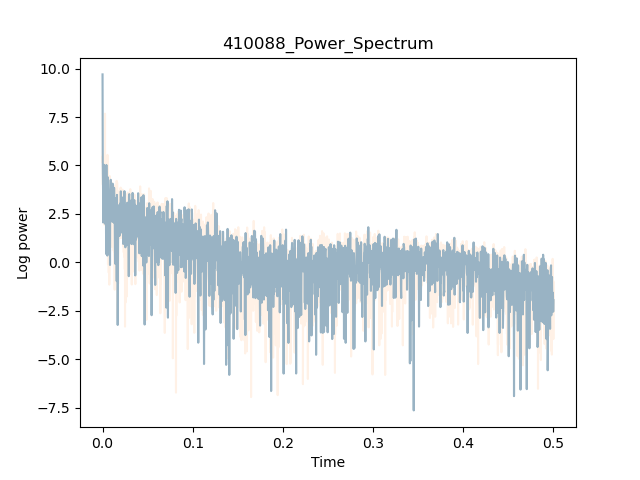}
        \caption{}
    \label{fig:Spectral_2}
    \end{subfigure}
    \begin{subfigure}[b]{0.48\textwidth}
        \includegraphics[width=\textwidth]{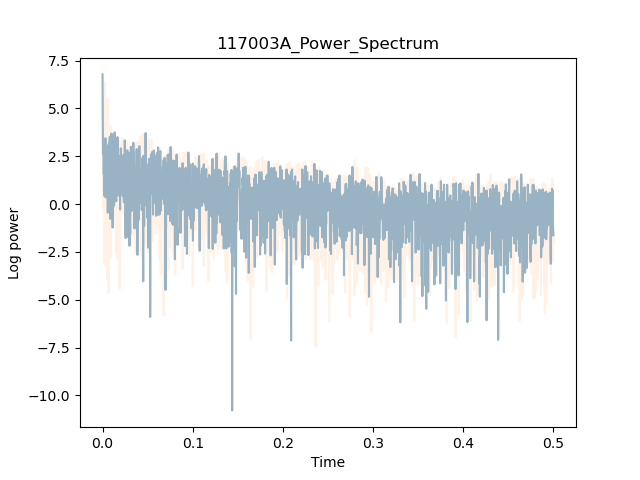}
        \caption{}
    \label{fig:Spectral_3}
    \end{subfigure}
    \begin{subfigure}[b]{0.48\textwidth}
        \includegraphics[width=\textwidth]{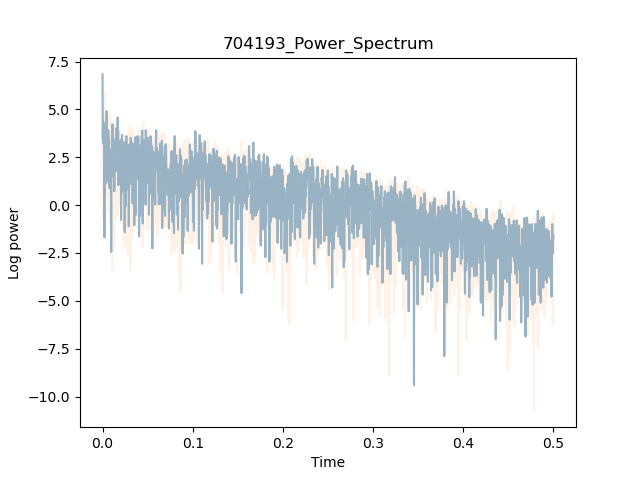}
        \caption{}
    \label{fig:Spectral_4}
    \end{subfigure}
    \caption{4 candidate flow time series' log power spectral densities from various locations in Australia. All power spectra appear to exhibit significant power at low frequencies, with the power eroding across higher frequency components.}
    \label{fig:Spectral_hydrological_time_series}
\end{figure*} 

Figure \ref{fig:Spectral_hydrological_time_series} shows the four power spectral density estimates corresponding to the underlying temporal processes in Figure \ref{fig:Time_hydrological_time_series}. Figures \ref{fig:Spectral_1}, \ref{fig:Spectral_2}, \ref{fig:Spectral_3} and \ref{fig:Spectral_4} display remarkable similarity. All figures exhibit a low frequency component which exhibits significant power, and subsequently lose power as we move toward higher frequency components. The low frequency component corresponding to significant power is approximately located at  $\nu \approx 1/365$, which corresponds to an annual periodic component - confirming the inherent seasonality in the data.  


Given the clear similarity in these stations' power spectra, we wish to further explore spectral similarity across the entire collection. We normalize for the amplitude of our power spectra with the same procedure as in Section \ref{Time_domain_analysis}. Let $\tilde{f}^{*}_i(\nu_j)= \frac{f^{*}_i(\nu_j)}{||f^{*}_i||}$ be our collection of optimal spectral trajectories. The vectors $\tilde{f}^{*}_i(\nu_j)$ reflect relative differences in various stations' spectra, and capture differences in key frequency components between streamflow spectra. 

Analogously to Section \ref{Time_domain_analysis}, we now define a \emph{Spectral distance matrix} $D^{S}_{ab}= \sum^{m-1}_{\nu_j=0} |\tilde{f}^{*}_a(\nu_j) - \tilde{f}^{*}_b(\nu_j)|$ that measures distance between  normalized spectral trajectories. To better interpret inherent collective similarity, we transform our distance matrix
\begin{equation}
A^{S}_{ab} = 1 - \frac{D_{ab}^{S}}{\max D^{S}},   
\end{equation}
and apply hierarchical clustering to the resulting \emph{spectral affinity matrix}. Again, this affinity matrix $A^{S} \in \mathbb{R}^{n \times n}$ is normalised so that all elements lie in $[0,1]$.
A high score (close to 1) in the affinity matrix $A^S$ would indicate that two hydrological stations exhibit similarly high power in key frequency components of their streamflow. A low score (close to 0) would indicate that two stations exhibit dissimilarity in their streamflow's key frequency components.  

\begin{figure*}
    \centering
    \includegraphics[width=0.95\textwidth]{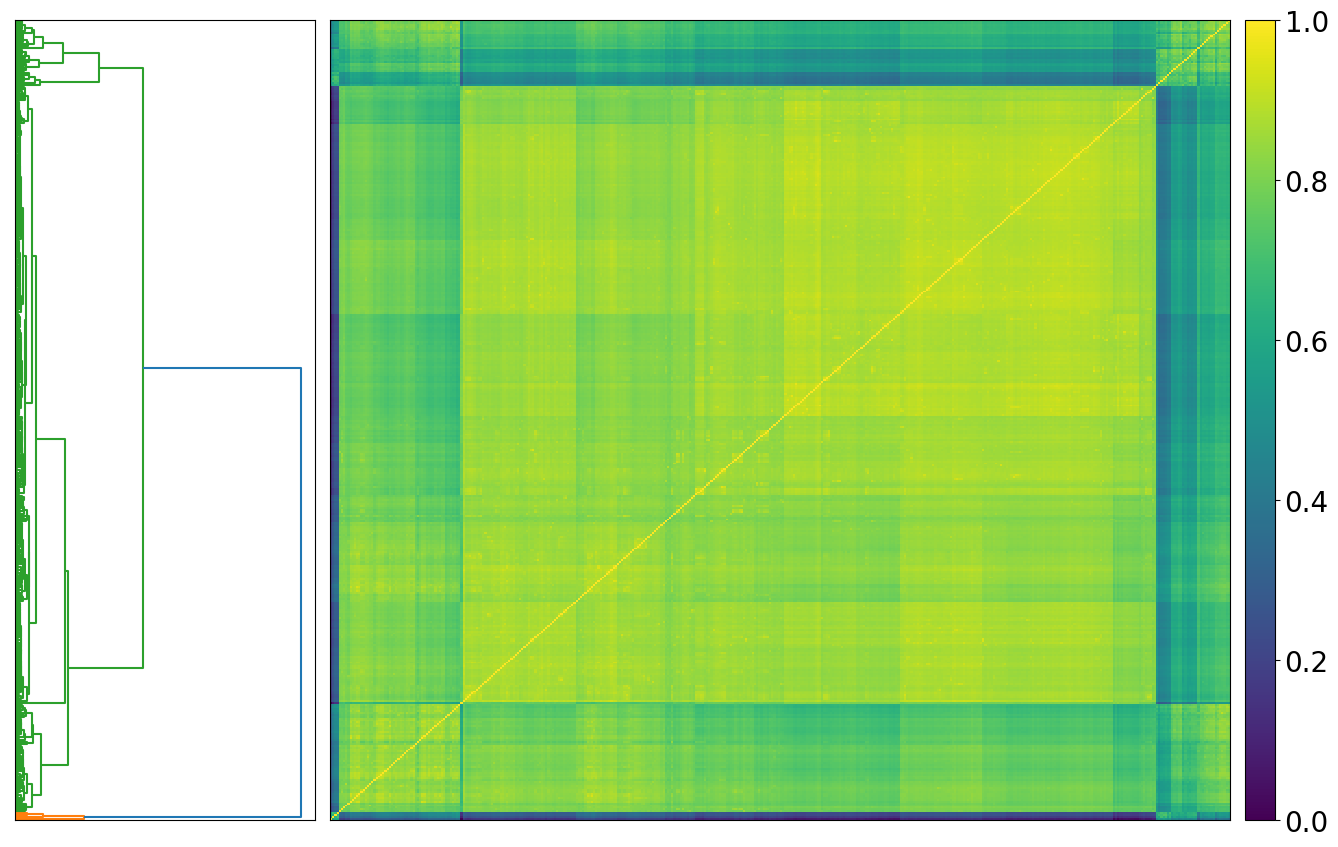}
    \caption{Spectral trajectory distance matrix applied to flow time series. There is one predominant cluster, and this highlights the overwhelming similarity in frequency behaviours among all streamflow time series. The similarity in periodic behaviour, relative to temporal behaviour, is striking. }
    \label{fig:Dendrogram_spectral_trajectories}
\end{figure*}

Figure \ref{fig:Dendrogram_spectral_trajectories} displays the hierarchical clustering results when applied to our spectral affinity matrix. The mean of our spectral affinity matrix, $ \nu^{S}$, is computed similarly (averaging the elements above the diagonal) and determined to be 0.79 - significantly higher than the average temporal similarity. Furthermore, the cluster structure reveals the overwhelming spectral similarity among the collection. There is one predominant cluster (consisting of a small sub-cluster), and an outlier cluster which consists of only a small subset of hydrology stations. This analysis confirms that the periodic dynamics of the streamflow time series are highly uniform across the entire collection - where almost all streamflow spectra exhibit a high power low frequency component, and incrementally less power as one examines the spectrum against higher frequency components. 



\section{Governing physical process}
\label{Governing_physical_process}
Having observed the pronounced temporal and spectral similarity exhibited by Australian hydrological stations, regardless of their spatial location, we hypothesize that there may be a governing physical process which captures the dynamics of Australian streamflow. In this section we formulate an overall process, with each time series representing a potentially shifted version of this process.

 For each of our time series $x_i(t)$, we seek to learn an offset $\phi_i$, where each time series has a unique offset. We bound each $\phi_i$ such that it lies in $\{0,...,365\}$. Let $\phi_1,...,\phi_n$ denote individual offsets for the time series $x_1,...,x_n$, and $G$ be an overall governing process. We seek the offsets and overall process, $G$, to be
\begin{equation}
\left (\hat{G}, \hat{\phi}_1, ..., \hat{\phi}_n \right ) = \argmin_{G, \phi_1,...,\phi_n} \sum^n_{i=1} \sum^{T-\phi_i}_{t=1} \bigg( x_i(t+\phi_i) - G(t) \bigg)^2
\end{equation}

To minimize this loss function, we perform a sequence of updating steps. First, we update the governing process  $G$. Here, we compute the mean of the current iteration of our governing process time series. Then, we update $\phi_1,...,\phi_n$. This step involves updating the alignments for the underlying stations' time series. Then, we update the governing process, $G$, by computing another mean. Next, update $\phi_1,...,\phi_n$. This procedure converges quickly after a few iterations. For initialization, we take $G (t) = \frac{1}{n} \sum_{i=1}^n x_i (t)$ to be the overall mean of the original, unshifted, time series.

\begin{figure*}
    \centering
    \includegraphics[width=0.95\textwidth]{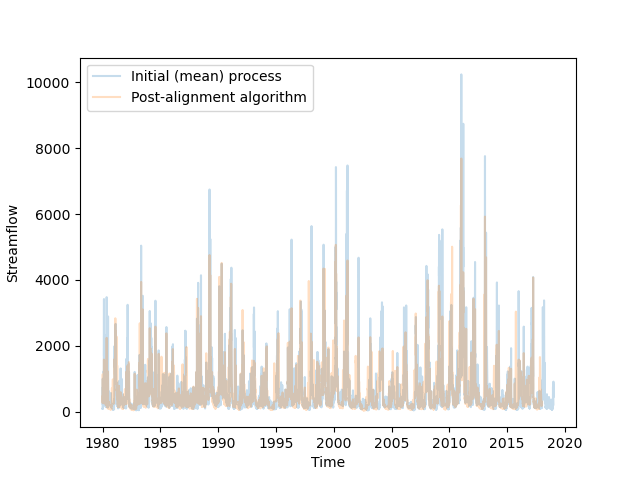}
    \caption{Temporal processes before and after our $L_2$ alignment procedure. The figure shows the initial process prior to the alignment algorithm and the post-alignment procedure. }
    \label{fig:Temporal_governing_process}
\end{figure*}

The initial estimate of the governing physical process and the estimate after the application of our algorithm  are shown in Figure \ref{fig:Temporal_governing_process}. The figure reveals several interesting takeaways. First, the spikes in aggregate aligned streamflow are highly periodic. First inspection suggests that a prominent spike in streamflow exists roughly once every year. This is certainly the most obvious pattern in the time series, with no clear insights in the periods between the big spikes. The peaks correspond to the significant power observed at low frequencies in our spectral analysis. 

\begin{figure*}
    \centering
    \begin{subfigure}[b]{0.48\textwidth}
        \includegraphics[width=\textwidth]{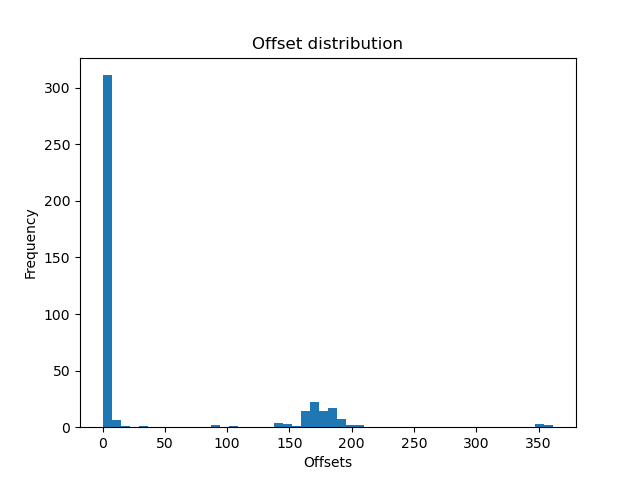}
        \caption{$\phi$ offsets}
    \label{fig:phi_offsets}
    \end{subfigure}
    \begin{subfigure}[b]{0.48\textwidth}
        \includegraphics[width=\textwidth]{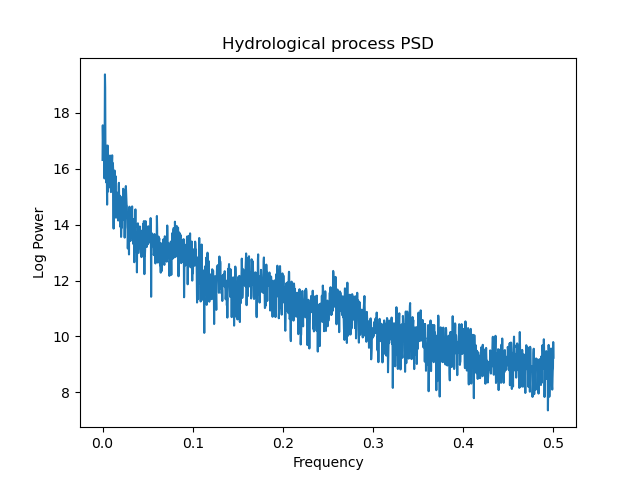}
        \caption{Governing spectral process}
    \label{fig:Spectral_governing_process}
    \end{subfigure}
    \caption{Distribution of $\phi$ offsets for underlying hydrological processes, and power spectral density of the governing physical process. One can see that the predominant number of offsets are distributed $\sim$ 0, and the governing physical process behaves similarly to most underlying power spectra.}
    \label{fig:Governing_physical_process_PSD_offsets}
\end{figure*}

The second key observation is the broad similarity between the initial mean process, and the determined governing physical process, which minimizes our loss function. The distribution of $\phi$ offsets, shown in figure \ref{fig:phi_offsets} confirms this. Approximately three quarters of the series are determined to have no offset or a minimal offset, suggesting that in most cases simply taking the mean of the underlying time series would serve as a good approximation to the overall process. An exhaustive list of non-zero offsets, separated by state, are shown in Table \ref{tab:State_offset_table}. The three states with the most significant number of non-zero offsets are Tasmania, Victoria and Western Australia. Interestingly, these three states also share the largest median (non-zero) offset value. However as a percentage of its total number of hydrology stations, Victoria's 44 stations with non-zero offsets is significantly lower than Tasmania, Western Australia and Tasmania. As a percentage of the number of stations, the states that are most likely to exhibit phase shifting dynamics in the time domain are Tasmania (81\%), Western Australia (69\%) and South Australia (46.2\%). As these states possess hydrology stations located at some of the most extreme points in Australia (by way of their spatial coordinates), this may indicate that there is some relationship between the offset $\phi$, and a candidate hydrology station's spatial location.


\begin{table}
\centering
\begin{tabular}{ |p{3.5cm}||p{2.8cm}|p{2.9cm}|p{1.2cm}|}
 \hline
 \multicolumn{4}{|c|}{\# non-zero offsets, \% non-zero, and median value by state } \\
 \hline
 State & \# non-zero offsets & \% non-zero offsets & $\mu^{\phi}$  \\
 \hline
 ACT & 0 & 0\% & n/a \\
 Northern Territory & 5 & 33.3\% & 87 \\
 NSW & 14 & 13.2\% & 80.5 \\
 Queensland & 11 & 13.75\% & 3 \\
 South Australia & 6 & 46.2\% & 11.5 \\
 Tasmania & 17 & 81\% & 181 \\
 Victoria & 44 & 36.6\% & 172 \\
 Western Australia & 36 & 69\% & 168 \\
\hline
\end{tabular}
\caption{Cardinality of each state's non-zero offsets and median non-zero offset value.}
\label{tab:State_offset_table}
\end{table}

Figure \ref{fig:Spectral_governing_process} shows the power spectral density for the governing physical process. Unsurprisingly, it displays the same key behaviours as the underlying stations' streamflow time series. First, there is a low frequency high power component, indicative of the annual periodicity in streamflow, and reduced power in higher frequency components.


\section{Evolutionary Dynamics in governing and underlying processes}
\label{Evolutionary_Dynamics}

Section \ref{Governing_physical_process} demonstrated a framework to determine a governing streamflow process. In doing so, we highlighted several key points. Critically, the underlying hydrological time series closely resemble the governing physical process' behaviours. Most notably the temporal dynamics share a periodic spike in streamflow, which corresponds to high power at low frequencies, and decreasing power at higher frequency components in the frequency domain. However, it is possible that temporal or spectral dynamics may change over time if the underlying process is nonstationary. Furthermore, the underlying time series and the governing process may display differences in their behaviour prior to and after our alignment algorithm. In this section, we test the existence of nonstationarity among our collection of underlying streamflow processes, and in the governing physical process before and after the procedure introduced in Section \ref{Governing_physical_process}.


First, we turn to our collection of underlying streamflow time series. We start by generating a sequence of rolling correlation matrices, indexed by time. We select a rolling window of $w=365$ days, and compute the correlation matrix
\begin{align}
    \Omega_{ij} (t) =\frac{\sum_{k=t-365}^t (x_i(k) - \bar{x}_i)(x_j(k) - \bar{x}_j))}{\left(\sum_{k=t-365}^t (x_i(k) - \bar{x}_i)^2 \sum_{k=t-365}^t (x_j(k) - \bar{x}_j)^2\right)^{1/2}}.
\end{align}
In the equation above, $\bar{x}_i$ denotes the mean of the hydrology time series $x_i(k)$ over any candidate rolling interval $k=t-365,...,t$. At each point in time, we compute an eigendecomposition and study the evolution of our eigenvalues $\lambda_1(t),...,\lambda_n(t)$ and eigenvectors $v_1(t),...,v_n(t)$. To explore the explanatory variance exhibited by our first eigenvalue, a measure of the degree of collective behaviour within our streamflow data, we compute a normalized first eigenvalue $\tilde{\lambda}_1(t) = \frac{\lambda_1(t)}{\sum^n_{j=1} \lambda^1_j (t)} $.

We analogously compute the rolling correlation matrix for our collection of aligned time series, $\Omega^{\phi}_{ij} (t)$, and let the time-varying vector of eigenvalues and eigenvectors of our aligned time series' be $\lambda^{\phi}_1(t),...,\lambda^{\phi}_n(t)$ and $v^{\phi}_1(t),...,v^{\phi}_n(t)$ respectively. We let the normalized first eigenvalue of our aligned collection be, $\tilde{\lambda}^{\phi}_1(t) = \frac{\lambda^{\phi}_1(t)}{\sum^n_{j=1} \lambda^{1, \phi}_j (t)}$.

\begin{figure*}
    \centering
    \begin{subfigure}[b]{0.48\textwidth}
        \includegraphics[width=\textwidth]{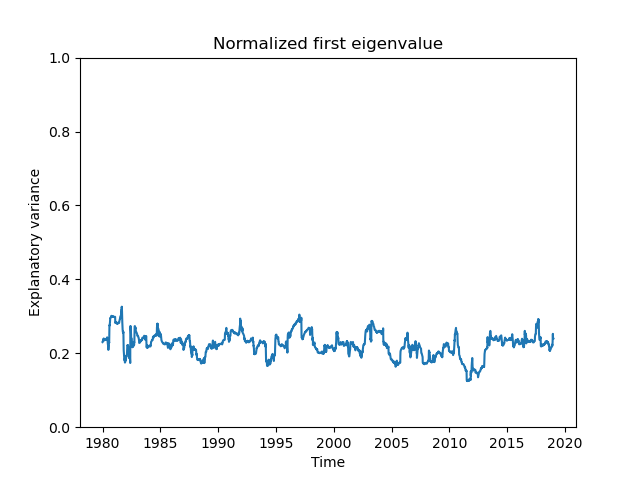}
        \caption{Initial collection}
    \label{fig:Eigenvalue_raw}
    \end{subfigure}
    \begin{subfigure}[b]{0.48\textwidth}
        \includegraphics[width=\textwidth]{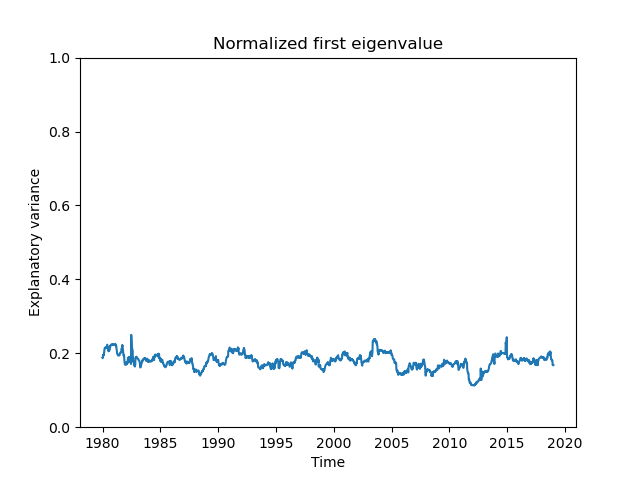}
        \caption{Aligned collection}
    \label{fig:Eigenvalue_aligned}
    \end{subfigure}
    \caption{Evolution of normalized first eigenvalue for raw and aligned collections of streamflow. Both the initial and aligned collection display broad similarity in the behaviour of the normalized first eigenvalue. }
    \label{fig:Eigenvalue_evolution}
\end{figure*} 

\begin{figure*}
    \centering
    \begin{subfigure}[b]{0.75\textwidth}
        \includegraphics[width=\textwidth]{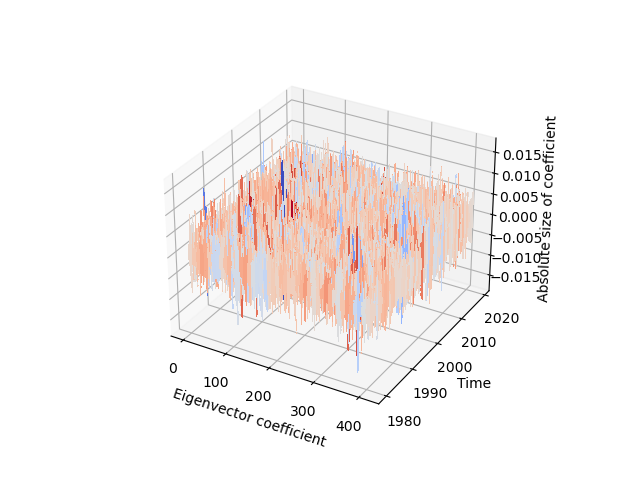}
        \caption{}
    \label{fig:Eigenvector_raw}
    \end{subfigure}
    \begin{subfigure}[b]{0.75\textwidth}
        \includegraphics[width=\textwidth]{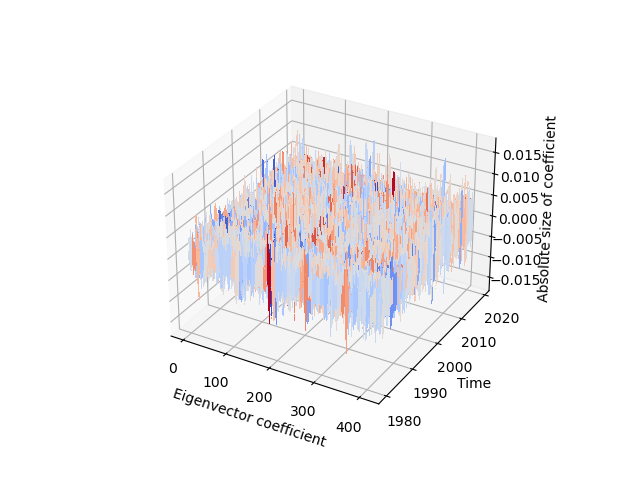}
        \caption{}
    \label{fig:Eigenvector_aligned}
    \end{subfigure}
    \caption{Evolution of first eignenvector magnitude of coefficients. The figure shows the magnitude of the coefficient of each element along the basis for each eigenvector propagated forward in time. }
    \label{fig:Eigenvector_evolution}
\end{figure*}

In Figure \ref{fig:Eigenvector_evolution} we show the normalized first eigenvalues, $\tilde{\lambda}_1(t)$ and $\tilde{\lambda}^{\phi}_1(t)$ for the initial and aligned time series collections respectively. There are some subtle differences between the functions. First, $\tilde{\lambda}_1(t)$ which is displayed in figure \ref{fig:Eigenvalue_raw}, exhibits more pronounced periodicity and a higher amplitude than $\tilde{\lambda}^{\phi}_1(t)$ which is shown in figure \ref{fig:Eigenvalue_aligned}. That is, the first eigenvalue in the initial (unaligned) case accounts for more explanatory variance than in the aligned case. At first, this may seem counter-intuitive. However, after aligning the time series as described in Section \ref{Governing_physical_process}, a significant portion of the variance in our collection has now been accounted for. So one could argue that the $\tilde{\lambda}^{\phi}_1(t)$ captures a different feature of the collection's variance, that which has not already been removed by alignment, and may provide similar insights to the explanatory variance exhibited by the initial collection's second eigenvalue. 

Next, we explore the evolutionary behaviour of both collections' first eigenvector. The evolutionary eigenvectors are shown in figure \ref{fig:Eigenvalue_evolution}, with figure \ref{fig:Eigenvector_raw} and figure \ref{fig:Eigenvector_aligned} showing the initial collection and aligned collection's first eigenvector, respectively. The first notable observation is the general uniformity in magnitude along all coefficients. That is, each hydrological station comprises a similar level of attribution to the first eigenvector's total variance. Furthermore, this trend is consistent over time in both the initial and aligned collections. The second observation is a moderate increase in the uniformity of magitudes in the aligned collection, shown in figure \ref{fig:Eigenvector_aligned}. This is consistent with the core aim of our procedure, which translates the underlying time series such that the $L^2$ norm with the current governing process, $G$, is minimized. To further explore this phenomenon, we compute the variance of time-varying eigenvector coefficients among both collections. The variance among the collection of initial time series is 1.03 x $10^{-5}$, while the aligned collection is 9.43 x $10^{-6}$. The modest reduction in variance is consistent with the increased uniformity observed in the figure. The reduction in variance of $\sim 7.5\%$ may indicate that the initial mean of the underlying time series, prior to the optimisation procedure, serves as a good estimate of an underlying physical process.



\begin{figure*}
    \centering
    \begin{subfigure}[b]{0.75\textwidth}
        \includegraphics[width=\textwidth]{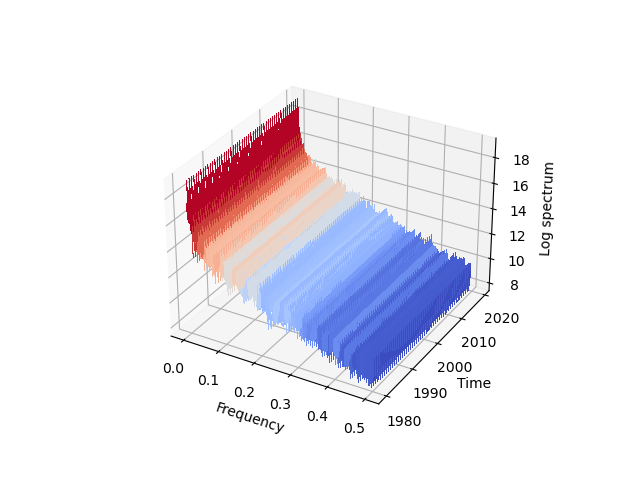}
        \caption{}
    \label{fig:Governing_process_initial}
    \end{subfigure}
    \begin{subfigure}[b]{0.75\textwidth}
        \includegraphics[width=\textwidth]{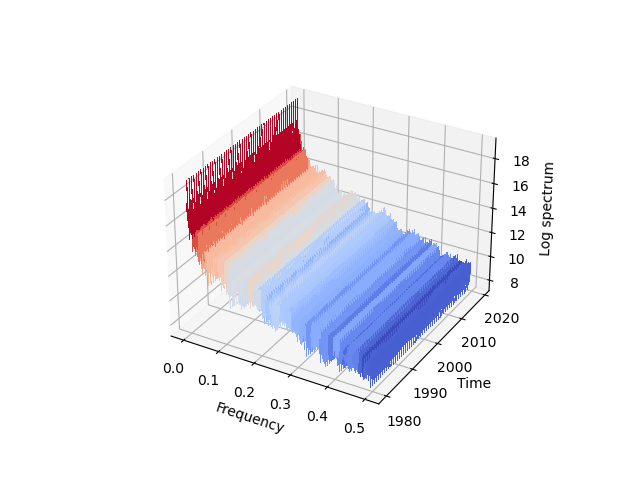}
        \caption{}
    \label{fig:Governing_process_aligned}
    \end{subfigure}
    \caption{Rolling log power spectral density of governing temporal process. The constant nature of the log PSD suggest that both spectra are stationary.}
    \label{fig:Evolutionary_dynamics_spectrum}
\end{figure*}

 
 In keeping with our assessment of time-varying dynamics, we now turn to analyzing the power spectrum of the governing hydrological process. We wish to answer two main questions. First, do the spectral dynamics of the overall process resemble the dynamics of the underlying time series? Second, do we see differences over time in the power spectrum? Figure \ref{fig:Evolutionary_dynamics_spectrum} shows the time-varying power spectrum of our governing process $G$ before and after our alignment procedure. 
 
 The spectral dynamics of the governing process display key similarities with the underlying collection of time series. Most importantly, they exhibit greatest power at the same frequency component, $\nu \approx 1/365$, representing the annual periodic component. Like the underlying time series, lower power is exhibited at higher frequency components. 

 Visual inspection of Figure \ref{fig:Evolutionary_dynamics_spectrum} suggests that the power spectrum does not display significant variability over time, for either collection (figure \ref{fig:Governing_process_initial} or figure \ref{fig:Governing_process_aligned}). At the beginning of our time window in 1980, the spectrum displays significant power at low frequencies, with decreasing power at higher frequencies. Although the precise amplitude at various frequencies does oscillate over time, the general pattern is relatively stable. The spectral similarity between the initial and aligned collections once again suggests that simply computing an average of the initial time series could serve as a good approximation to the underlying physical process.

\section{Conclusion}
\label{conclusion}

In Section \ref{Time_domain_analysis} we studied the similarity between hydrological stations' temporal dynamics. Our trajectory analysis and associated hierarchical clustering revealed the existence of three predominant clusters. The three clusters shown in Figure \ref{fig:Dendrogram_temporal_trajectories} display a high level of intra-cluster similarity, with surprisingly low inter-cluster similarity. The temporal affinity norm of 0.26 highlights the low degree of affinity exhibited between trajectories belonging to separate clusters. Subsequent experiments reveal that these varied cluster behaviours were not related to stations' spatial location, ruling out the assertion that the temporal dynamics of Australian streamflow may be related to geographic location. Despite the clear cluster separation, visual inspection suggests that most streamflow time series possess broadly similar features - small oscillations around a level close to zero, with sharp spikes which appear to occur on an annual periodicity.

Section \ref{Spectral_domain_analysis} builds upon this analysis, introducing a new framework to explore similarity in periodic behaviours. Our experiments reveal uniform behaviours in the spectral dynamics of underlying stations' streamflow. Figure \ref{fig:Dendrogram_spectral_trajectories} shows the existence of a majority cluster of spectral trajectories, and the spectral affinity norm of 0.79 confirms the profound similarity in streamflow spectral dynamics. This is primarily due virtually all locations displaying most significant power at at the frequency $ \nu = 0.0027 $, corresponding to an annual streamflow cycle. 

Given the general similarity in the temporal domain and the profound similarity displayed in the frequency domain, in Section \ref{Governing_physical_process} we introduce an iterative algorithmic procedure to determine a governing hydrological process. We introduce an $L^2$ based objective function, where each time series is translated to minimize an average loss. Figures \ref{fig:Temporal_governing_process} and \ref{fig:Governing_physical_process_PSD_offsets} reveal several notable findings. First, the mean of all streamflow time series does a reasonably good job in estimating the general behaviour of the collection, and is similar to our estimated process after the application of our alignment procedure. Figure \ref{fig:phi_offsets} shows that the majority of estimated offsets are in fact 0, confirming the sample mean's utility as a rough estimate. Figure \ref{fig:Spectral_governing_process} demonstrates that the power spectrum of the governing process resembles that of the underlying time series, and therefore still resembles the overarching theme of periodic streamflow behaviour.

We round out our analysis in Section \ref{Evolutionary_Dynamics} where we examine the stationarity of such behaviours previously identified in the prior sections. In our time-varying PCA, we study the explanatory variance exhibited by the correlation matrix's first eigenvalue, which is shown in Figure \ref{fig:Eigenvalue_evolution}. The figure demonstrates a similar level of explanatory variance exhibited by the first eigenvalue in both the initial and aligned collections of streamflow time series. To further investigate, we accompany this analysis with a time-varying study of the magnitude of the first eigenvectors' coefficients displayed by both collections. Both collections display relative uniformity in the coefficient magnitudes over time, indicating that there is no element, or collection of elements, along the vector accounting for the majority of the variance. Figure \ref{fig:Evolutionary_dynamics_spectrum} shows the evolutionary power spectral density for both collections. It is clear that in both cases, the spectrum displays pronounced stationarity over time, exhibiting the same key features of significant power at low frequencies and decreasing power along higher frequency components.  

There are several avenues for future work. First, one could explicitly compare the offset methodologies introduced in this paper to other techniques used in various areas of the physical sciences. Furthermore, one could study the sensitivity of the results as various objective functions are used instead of those introduced in our experiments. The spectral analysis techniques introduced in this manuscript could be implemented differently, specifically, one could do so in a Bayesian framework so that uncertainty is quantified when evaluating the power spectrum. Although many of the techniques used in this paper are relatively standard (spectral analysis, $L_1$ normalized trajectory modelling, hierarchical clustering, PCA, etc.) there are alternative methods that could be used to address the same data-driven questions. Future work could explore the stability of our results when alternative techniques are used. 

\section{Appendix}
\label{Appendix}

\subsection{Spatial dynamics}
\label{spatial_dynamics}

In section \ref{Time_domain_analysis} and section \ref{Spectral_domain_analysis} of the paper, four hydrological stations' time series are displayed (Paddy's River, Goodradigbee, Bluewater Creek and Gascoyne). The first two correspond to hydrology stations on the East Coast of Australia, while the third and fourth correspond to stations in the Northern and Western parts of Australia respectively. Our analysis throughout the paper demonstrates that these time series all display high similarity in temporal and spectral dynamics. Given the geographic spread of such stations, this inspired us to explore the spatial distribution of hydrology stations across Australia.

\begin{figure*}
    \centering
    \begin{subfigure}[b]{0.48\textwidth}
        \includegraphics[width=\textwidth]{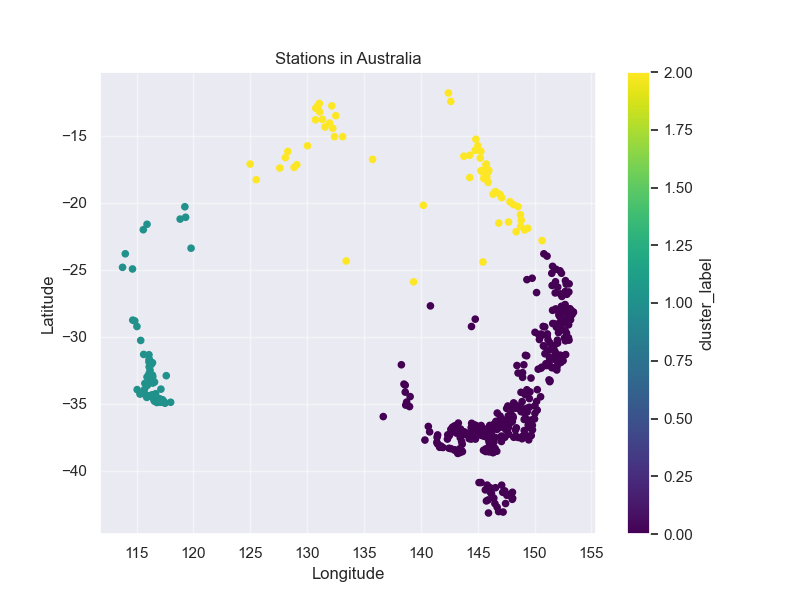}
        \caption{}
    \label{fig:Hydrology_map}
    \end{subfigure}
    \begin{subfigure}[b]{0.48\textwidth}
        \includegraphics[width=\textwidth]{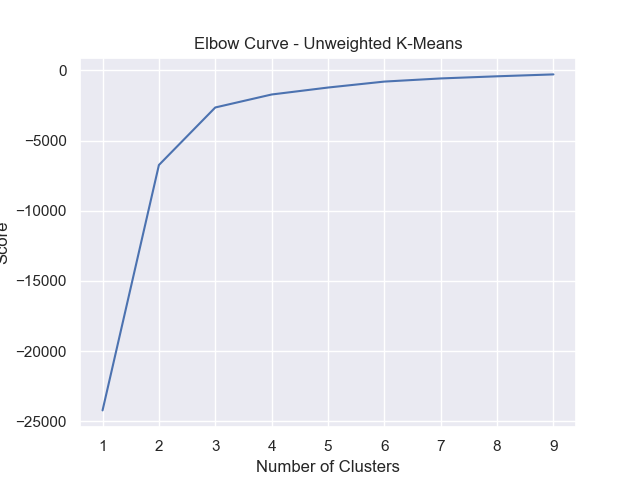}
        \caption{}
    \label{fig:Spatial_elbow}
    \end{subfigure}
    \caption{Location of hydrological stations annotated on map of Australia. Colours denote cluster labels, which are determined solely based on geodesic distance between hydrological stations.}
    \label{fig:Spatial_hydrological_plots}
\end{figure*} 

We generate a distance matrix between all $n=414$ hydrology stations, where the distance between any two candidate stations is the geodesic distance between two tuples of latitude and longitude data. We then apply \emph{K-means clustering} to determine hydrology stations' cluster memberships, where we use the elbow method to determine the existence of $K=3$ clusters. 

The algorithm's objective function seeks to minimize an appropriate sum of square distances. With $k$ chosen \textit{a priori} based on the elbow method aforementioned, we examine all possible disjoint unions $C_1 \cup C_2 \cup \dots \cup C_k $ of $\{ x_1,\dots, x_n \}$. Let $z_j$ be the \emph{centroid} of the subset $C_j$. One then tries to minimize the sum of square distances within each cluster to its candidate centroid:
$$ \sum_{j=1}^k \sum_{x \in C_j} ||x - z_j||^2 $$
For a space of dimension 2 or greater, it is NP-hard to find the globally optimal solution of this problem. The K-means algorithm  \cite{Lloyd1982} is an iterative solution that enables fast convergence to a locally optimal solution.

Figure \ref{fig:Spatial_hydrological_plots} shows the two respective sub-figures relating to our cluster analysis. Figure \ref{fig:Hydrology_map} corresponds to the geographic location of hydrology stations (with their colour denoting cluster membership) and the elbow curve (shown in Figure \ref{fig:Spatial_elbow} displays the elbow curve plot. One can clearly identify three separate geographic clusters in Figure \ref{fig:Hydrology_map}. However the preceding analysis in this paper demonstrated that despite clear geographic separation, temporal and spectral dynamics of such stations were largely indistinguishable.

\subsection{Mathematical objects}
\label{mathematical_objects}

\begin{table}[H]
\begin{tabular}{ |p{2.3cm}||p{8.9cm}|}
 \hline
 \multicolumn{2}{|c|}{\textbf{Mathematical objects table}} \\
 \hline
 Object & Description \\
 \hline
 $n$ & \# hydrology stations \\ 
 $x_i(t)$ & Streamflow time series \\ 
 $\tilde{x}_i(t)$ & Streamflow trajectory time series \\
 $D^{T}$ & Temporal distance matrix \\
 $A^{T}$ & Temporal affinity matrix \\
 $f(\nu_{k})$ & Latent power spectral density \\
 $I(\nu_{j})$ & Periodogram estimate of power spectrum \\
 $f_i^{W}(\nu_{j})$ & Welch's estimate of power spectral density \\
 $f_i^{*}(\nu_j)$ & Optimal spectra time series \\
 $S^{*}$ & Optimal size of each segment \\
 $\omega^{*}$ & Optimal overlap between neighbouring segments \\
 $\tilde{f}^{*}_i(\nu_j)$ & Optimal spectral trajectory time series \\
 $D^{S}$ & Spectral distance matrix \\
 $A^{S}$ & Spectral affinity matrix \\
 $\phi_i$ & Offset for streamflow time series $i$ \\
 $G$ & Governing streamflow process \\
 $\Omega(t)$ & Initial collection time-varying correlation matrix \\
 $\lambda_i(t)$ &  Initial collection time-varying eigenvalues \\
 $v_i(t)$ & Initial collection time-varying eigenvectors \\
 $\tilde{\lambda}_1(t)$ & Initial collection time-varying normalized first eigenvalue \\
 $\Omega^{\phi}(t)$ & Aligned collection time-varying correlation matrix \\
 $\lambda^{\phi}_i(t)$ &  Aligned collection time-varying eigenvalues \\
 $v^{\phi}_i(t)$ & Aligned collection time-varying eigenvectors \\
 $\tilde{\lambda}^{\phi}_1(t)$ & Aligned collection time-varying normalized first eigenvalue \\
 \hline
\end{tabular}
\caption{Mathematical objects and definitions}
\label{tab:MathematicalObjects1}
\end{table}

\subsection{Nonstationarity analysis}
\label{Nonstationary_analysis}

To further explore the nonstationarity in the initial and aligned governing processes, we apply a \emph{reversible jump Markov Chain Monte Carlo} algorithm (AdaptSPEC \cite{Rosen2012}) to identify the number and location of change points in each time series. The algorithm determines changepoints based on changes in the time-varying power spectrum for any candidate time series. That is, a nonstationary time series is assumed to exhibit local stationarity, where each (stationary) partition possesses a unique power spectral density. The algorithm determined the existence of 9 changepoints in both the initial and aligned time series. However, the location of change points within each time series were quite different. Changepoint indices and the respective location (date) for both time series are shown in Table \ref{tab:table_changepoints}.


\begin{figure*}
    \centering
    \begin{subfigure}[b]{0.75\textwidth}
        \includegraphics[width=\textwidth]{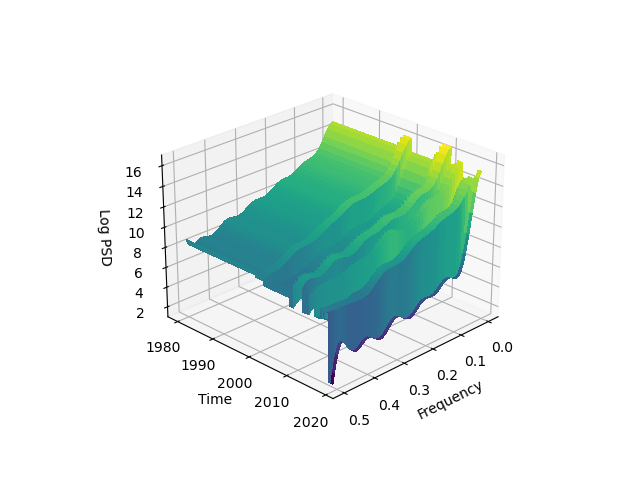}
        \caption{}
    \label{fig:Initial_surface}
    \end{subfigure}
    \begin{subfigure}[b]{0.75\textwidth}
        \includegraphics[width=\textwidth]{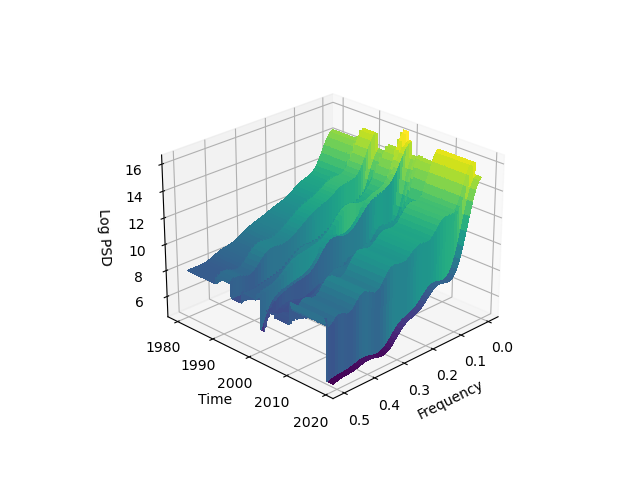}
        \caption{}
    \label{fig:Aligned_surface}
    \end{subfigure}
    \caption{Time-varying power spectral density estimation for initial and aligned collections of hydrological time series.}
    \label{fig:Time_varying_spectra_adaptspec}
\end{figure*} 

\begin{table}[h]
\begin{center}
\begin{tabular}{ |p{1.8cm}||p{2.75cm}|p{3.5cm}|}
 \hline
 \multicolumn{3}{|c|}{Changepoints and changepoint location} \\
 \hline
 Collection & Changepoint index & Changepoint location (MAP estimate)  \\
 \hline
 Initial & 1/9 & 28/1/99 \\
 Initial & 2/9 & 27/3/01 \\
 Initial & 3/9 & 23/11/07 \\
 Initial & 4/9 & 20/01/09 \\
 Initial & 5/9 & 08/04/11 \\
 Initial & 6/9 & 01/01/13 \\
 Initial & 7/9 & 20/04/14 \\
 Initial & 8/9 & 22/09/15 \\
 Initial & 9/9 & 20/12/17 \\
 Aligned & 1/9 & 01/03/88 \\
 Aligned & 2/9 & 13/07/92 \\
 Aligned & 3/9 & 16/6/95 \\
 Aligned & 4/9 & 9/8/99 \\
 Aligned & 5/9 & 14/10/00 \\
 Aligned & 6/9 & 23/10/01 \\
 Aligned & 7/9 & 06/08/04 \\
 Aligned & 8/9 & 6/08/08 \\
 Aligned & 9/9 & 11/04/18 \\
\hline
\end{tabular}
\caption{AdaptSPEC changepoint indices and locations}
\label{tab:table_changepoints}
\end{center}
\end{table}

Figure \ref{fig:Time_varying_spectra_adaptspec} shows the time-varying spectral densities for the initial and aligned time series collections. In both instances, all partitions exhibit a power spectrum with high similarity in key trends. That is, significant power at low frequencies (indicative of a strong annual periodicity), and decreasing power at higher frequency components. Although 10 (quite different) segments are identified within each time series, the consistency in the shape and amplitude of the power spectrum between all adjacent segments supports the general notion of stationarity for both time series.


\bibliographystyle{elsarticle-num-names}
\bibliography{__References.bib}
\end{document}